\documentclass[10pt]{article}
\usepackage{graphicx}

\def\Title#1{\begin{center} {\Large #1 } \end{center}}
\def\Author#1{\begin{center}{ \sc #1} \end{center}}
\def\Address#1{\begin{center}{ \it #1} \end{center}}

\newcommand\pubblock{\rightline{\begin{tabular}{l} Proceedings of the Fifth Annual LHCP\\ \pubnumber\\
         \pubdate  \end{tabular}}}

\newenvironment{Abstract}{\begin{quotation} \begin{center} 
             \large ABSTRACT \end{center}\bigskip 
      \begin{center}\begin{large}}{\end{large}\end{center} \end{quotation}}

\newenvironment{Presented}{\begin{quotation} \begin{center} 
             PRESENTED AT\end{center}\bigskip 
      \begin{center}\begin{large}}{\end{large}\end{center} \end{quotation}}

\def\beq{\begin{equation}}
\def\eeq#1{\label{#1}\end{equation}}
\def\eeqn{\end{equation}}

\def\beqa{\begin{eqnarray}}
\def\eeqa#1{\label{#1}\end{eqnarray}}
\def\eeqan{\end{eqnarray}}

\let\bar=\overbar

\def\Dslash{\not{\hbox{\kern-4pt $D$}}}
\def\dslash{\not{\hbox{\kern-2pt $\del$}}}

\def\msb{{\bar{\ssstyle M \kern -1pt S}}}

\usepackage{color}
\usepackage{amsmath}
\usepackage{cite}

\newcommand\pubnumber{TTP17-029}

\newcommand\pubdate{\today}

\def\affiliation{Institut f{\"u}r Kernphysik, Karlsruhe Institute of Technology,\\
  Hermann-von-Helmholtz-Platz 1,
  D-76344 Eggenstein-Leopoldshafen, Germany\vspace{1mm}\\
 Institut f{\"u}r Theoretische Teilchenphysik,
  Karlsruhe Institute of Technology, \\
Engesserstra\ss e 7,
  D-76128 Karlsruhe, Germany}

\begin{document}

\large
\begin{titlepage}
\pubblock

\vfill
\Title{Theoretical implications of recent heavy flavour measurements at the LHC}
\vfill

\Author{Monika Blanke}
\Address{\affiliation}
\vfill
\begin{Abstract}
\begin{minipage}{14cm}
Recent measurements have revealed a number of intriguing deviations from the Standard Model predictions in $B$ meson decays, in particular in observables testing lepton flavour universality. We review their experimental status and theoretical description in terms of effective Hamiltonians. We also discuss possible new physics interpretations in terms of simplified models and summarise their status in view of the stringent constraints from flavour physics and high-$p_T$ collider searches.  
\end{minipage}
\end{Abstract}
\vfill

\begin{Presented}
The Fifth Annual Conference\\
 on Large Hadron Collider Physics \\
Shanghai Jiao Tong University, Shanghai, China\\ 
May 15-20, 2017
\end{Presented}
\vfill
\end{titlepage}
\def\thefootnote{\fnsymbol{footnote}}
\setcounter{footnote}{0}
%

\normalsize 


\section{Introduction}

The Standard Model (SM) of particle physics not only is a self-consistent theory, but it also describes the experimental results obtained in collider experiments with an astonishing accuracy. Nonetheless, the presence of New Physics (NP) beyond the SM is required to address some open conceptual and astrophysical/cosmological problems. The dynamics behind electroweak (EW) symmetry breaking and the naturalness problem of the EW scale call for new degrees of freedom not much beyond the TeV scale. Similarly, the flavour sector with its many parameters and its very hierarchical structure suggests the existence of a more fundamental theory of flavour. On the astrophysical and cosmological side, neither the baryon asymmetry of the universe nor the observed dark matter and dark energy densities can be accounted for within the SM. 

While we thus have a number of good reasons to believe in NP and expect it to be in the reach of current experiments like the Large Hadron Collider (LHC), so far we have failed to observe any direct manifestation of its presence. The direct searches for new heavy particles at the LHC leave us with the puzzling insight that the new states are either heavier than we had thought, or that they are well hidden from the search channels studied so far. 
 
Nonetheless, over the recent years some intriguing anomalies emerged in the data on semi-leptonic and rare $B$ meson decays. The measured values of 
\begin{equation}\label{eq:RD}
R(D^{(*)}) = \frac{\mathcal{B}(B\to D^{(*)}\tau\nu)}{\mathcal{B}(B\to D^{(*)}\ell\nu)}\,,
\end{equation}
where $\ell=e,\mu$, deviate by $3.9\sigma$ from their SM predictions, hinting for the presence of new sources of lepton flavour universality (LFU) violation. In addition, a number of anomalies have emerged in the $b\to s\ell^+\ell^-$ transitions, with the most recent surprise of finding the ratio
\begin{equation}\label{eq:RK}
R_{K^*} = \frac{\mathcal{B}(B\to K^*\mu^+\mu^-)}{\mathcal{B}(B\to K^*e^+e^-) }
\end{equation}
more than $2\sigma$ below its extremely clean SM prediction. Again, LFU appears to be violated by NP. Even more intriguingly, the various anomalies in the $b\to s\ell^+\ell^-$ sector can all be consistently resolved by a common NP contribution.

This letter provides a review of the status and recent developments in the theoretical interpretation of the anomalies described above. It is based on an overview talk given at the LHCP2017 conference in Shanghai. Section \ref{sec:RD} is devoted to the discussion of the $R(D^{(*)})$ anomalies and possible NP explanations. In section \ref{sec:bsmm} we turn our attention to the $b\to s\ell^+\ell^-$ transitions, where several anomalies have been observed in recent data. Section \ref{sec:sum} provides a summary and a brief outlook.

\section{\boldmath The semi-tauonic decays $B\to D^{(*)}\tau\nu$}
 \label{sec:RD}

The ratios $R(D)$ and $R(D^*)$, as defined in \eqref{eq:RD}, provide a test of LFU in semileptonic $B$ decays. As in the SM LFU is violated by the charged lepton masses $m_\tau \gg m_e,m_\mu$, the SM predictions \cite{Bernlochner:2017jka} for $R(D)$ and $R(D^*)$ differ from unity:
\begin{equation}
R(D)_\text{SM} = 0.299 \pm 0.003\,,\qquad
R(D^*)_\text{SM} = 0.257 \pm 0.003\,.
\end{equation}
The high precision is reached since the hadronic uncertainties largely cancel in the ratio.

Over the past years, BaBar \cite{Lees:2013uzd}, Belle \cite{Huschle:2015rga,Sato:2016svk,Hirose:2016wfn} and LHCb \cite{Aaij:2015yra,LHCb-new} presented a number of measurements of these ratios, yielding a consistent HFLAV fit \cite{Amhis:2016xyh}
\begin{equation}
R(D)_\text{exp} = 0.4074 \pm 0.046\,, \qquad
R(D^*)_\text{exp} = 0.305 \pm 0.015\,,
\end{equation} 
which exhibits a $3.9\sigma$ deviation from the SM, thereby hinting for new sources of LFU violation.

In the SM, the relevant $b\to c \tau\nu$ transition is mediated by a tree-level exchange of the $W^\pm$ boson. Model-independently, this transition can be described by the effective Hamiltonian
\begin{equation}\label{eq:bctaunu}
\mathcal{H}_\text{eff}^{b\to c\tau\nu} = \frac{4 G_F}{\sqrt{2}} V_{cb} \sum_j C_j \mathcal{O}_j + h.c.\,,
\end{equation}
with the operators
\begin{eqnarray}
\mathcal{O}_{V_{L,R}} &=& (\bar c\gamma^\mu P_{L,R} b)(\bar\tau\gamma_\mu P_L \nu)\,,\\
\mathcal{O}_{S_{L,R}} &=& (\bar c P_{L,R} b)(\bar\tau P_L \nu)\,,\\
\mathcal{O}_T &=& (\bar c\sigma^{\mu\nu} P_L b)(\bar\tau\sigma_{\mu\nu}P_L \nu)\,.
\end{eqnarray}
In the SM, only the Wilson coefficient $C_{V_L}=1$ is different from zero. 

Possible NP scenarios addressing the $R(D^{(*)})$ anomaly are the tree-level exchange of a new charged scalar particle \cite{Crivellin:2012ye,Crivellin:2013wna}, as arises in two Higgs doublet models, a heavy charged vector resonance $W'$ \cite{Greljo:2015mma}, or a scalar or vector leptoquark \cite{Fajfer:2015ycq,Becirevic:2016yqi}. A charged Higgs contribution would manifest itself via $C_{S_L}^\text{NP},C_{S,R}^\text{NP} \ne 0$, while for a $W'$ exchange $C_{V_L}^\text{NP} \ne 0$ or $C_{V_R}^\text{NP} \ne 0$, depending of the chirality of the new gauge interactions. Finally, in leptoquark models, different combinations of Wilson coefficients are generated, depending on the spin and gauge quantum numbers of the assumed leptoquark.

Global fits \cite{Fajfer:2012jt,Freytsis:2015qca,Bardhan:2016uhr} of the effective Hamiltonian \eqref{eq:bctaunu} to the data show that the $R(D^{(*)})$ anomaly can be resolved by the presence of NP in either the scalar or the vector Wilson coefficients. In the first case, $C_{S_L}^\text{NP}\simeq -C_{S_R}^\text{NP}$ has be fulfilled in order to generate NP effects of similar size in $R(D)$ and $R(D^*)$. In other words, the charged Higgs coupling to quarks has to be approximately pseudoscalar. In the second case, a good fit is obtained for $C_{V_L}^\text{NP}\ne
0$, i.\,e.\ for a $W'$ coupling to left-handed fermions. In both cases, the required NP contribution is quite large, having to compete with a tree-level process in the SM. 

Consequently, NP explanations of the $R(D^{(*)})$ anomaly are stringently constrained by complementary measurements in the flavour sector, but also by high-$p_T$ observables. A resolution of the anomaly by the scalar Wilson coefficients $C_{S_L}^\text{NP}\simeq -C_{S_R}^\text{NP}$ is challenged by the total $B_c$ lifetime \cite{Alonso:2016oyd}: The large pseudoscalar contribution required to reconcile $R(D^*)$ with the data generates a large contribution to the $B_c\to\tau\nu$ decay, due to the absence of a chirality suppression of (pseudo)scalar contributions. Scalar contributions to the $b\to c\tau\nu$ transition \cite{Celis:2016azn} also modify the $B\to D^{(*)}\tau\nu$ differential decay rates with respect to the SM. While the experimental precision is so far limited, the good agreement of the $B\to D\tau\nu$ differential rate with the SM prediction \cite{Lees:2013uzd,Huschle:2015rga} provides another hint against scalar contributions being the origin of the $R(D^{(*)})$ anomaly. 

A NP contribution to $C_{V_L}$, on the other hand, is not affected by the above constraints. Its contribution on the $B_c\to\tau\nu$ decay rate receives the same chirality suppression factor $m_\tau^2/m_b^2$ as in the SM and is therefore safely small. In addition, the differential decay rates remain the same as in the SM, as only the overall normalisation changes with the size of $C_{V_L}$. However it has been shown \cite{Feruglio:2016gvd,Feruglio:2017rjo} that loop diagrams involving  $C_{V_L}^\text{NP}$  generate deviations from the SM in $Z$ and $\tau$ decays in conflict with the data.

It hence appears that leptoquarks currently provide the best NP explanation for the $R(D^{(*)})$ anomaly.
However, also these models face stringent constraints. The  $(\bar c b)(\bar \tau\nu)$ operators\footnote{We suppress the Dirac structure for simplicity.}  are related by the $SU(2)_L$ gauge symmetry to operators like $(\bar bb)(\bar \tau\tau)$, $(\bar cc)(\bar \tau\tau)$, $(\bar sb)(\bar\tau\tau)$, and $(\bar sb)(\bar\nu\nu)$. The latter two are constrained by the upper bounds on the branching ratios of $B_s\to\tau^+\tau^-$ and $B\to K^{(*)}\nu\bar\nu$ \cite{Calibbi:2015kma,Crivellin:2017zlb}. The former two, on the other hand, are generated by the CKM mixing and lead to deviations from the SM in $\tau$ pair production at the LHC \cite{Faroughy:2016osc}, with the available data excluding major regions of the parameter space. The same interactions also mediate the decays $\Upsilon\to\tau^+\tau^-$ and $\psi\to\tau^+\tau^-$ \cite{Aloni:2017eny}.

In summary it thus turns out that a NP resolution of the $R(D^{(*)})$ anomaly is challenged by complementary, stringent constraints from $B$ decay observables, but also from EW precision measurements and high-$p_T$ searches. This is not unexpected, given that the required NP effect has to compete with a tree-level contribution in the SM and is therefore rather large.

\section{\boldmath The $b\to s\mu^+\mu^-$ transitions and $R_{K^{(*)}}$}
\label{sec:bsmm}

Let us now turn to another set of $B$ decay anomalies  that has recently attracted a lot of attention, related to the semileptonic $b\to s\mu^+\mu^-$ transition. The first indication of a deviation from the SM in this sector was provided by LHCb in 2013 \cite{Aaij:2013qta}\footnote{Note that the earlier hints for a non-standard forward-backward asymmetry $A_\text{FB}$ \cite{Wei:2009zv} were not confirmed by LHCb.}, observing a $3.7\sigma$ deviation from the SM in the angular observable $P'_5$ of the decay $B\to K^*\mu^+\mu^-$. This anomaly has been confirmed with more statistics by LHCb \cite{Aaij:2015oid}, while the precision achieved at Belle \cite{Wehle:2016yoi}, CMS \cite{CMS:2017ivg}, and ATLAS \cite{ATLAS:2017dlm} is currently too low to draw definite conclusions. A departure from the SM has also been found in the differential branching fraction of $B_s\to\phi\mu^+\mu^-$ \cite{Aaij:2015esa}.
Theoretically even more appealing are the substantial departures from unity found in the LFU ratios $R_K$ \cite{Aaij:2014ora} and $R_{K^*}$ \cite{Aaij:2017vbb}, as defined in \eqref{eq:RK}, as those are theoretically extremely clean.

The semileptonic $b\to s\mu^+\mu^-$ and radiative $b\to s\gamma$ transitions are conveniently described by the effective Hamiltonian
\begin{equation}
\mathcal{H}_\text{eff}= -\frac{4 G_F}{\sqrt{2}} V_{tb}^* V_{ts} \frac{e^2}{16\pi^2}\sum_i(C_i {\cal O_i} +C'_i {\cal O'_i})+h.c.\,,
\end{equation}
where the operators most sensitive to NP are the dipole operators 
\begin{equation}\label{eq:C7}
\mathcal{O}^{(\prime)}_7 =\frac{m_b}{e} (\bar s \sigma_{\mu\nu} P_{R(L)}b) F^{\mu\nu}
\end{equation}
and the four fermion operators 
\begin{equation}\label{eq:C910}
\mathcal{O}^{(\prime)}_9 = (\bar s\gamma_\mu P_{L(R)} b)(\bar\mu\gamma^\mu\mu)\,,\qquad
\mathcal{O}^{(\prime)}_{10}  = (\bar s\gamma_\mu P_{L(R)} b)(\bar\mu\gamma^\mu\gamma_5\mu)\,,
\end{equation}
that are not affected by tree-level contributions in the SM. Potential NP contributions therefore compete with the loop- and GIM-suppressed SM contribution, making a sizeable deviation from the latter much easier to achieve. Note that we neglect the scalar and pseudoscalar operators $\mathcal{O}^{(\prime)}_{S,P}$ in our discussion, as they are strongly constrained by the measured $B_s\to\mu^+\mu^-$ branching ratio \cite{Aaij:2017vad}, which is found in good agreement with the SM prediction \cite{Bobeth:2013uxa}.

The Wilson coefficients $C^{(\prime)}_{7,9,10}$ can be constrained by the measurement of various observables in radiative and semileptonic $b\to s$ transitions. For instance, the radiative decays $B\to X_s\gamma$, $B\to K^*\gamma$ etc. are governed only to the magnetic dipole operators $\mathcal{O}^{(\prime)}_7$, while semileptonic decays like $B\to K^{(*)}\mu^+\mu^-$, $B_s\to\phi\mu^+\mu^-$ and $B\to X_s\mu^+\mu^-$ are sensitive to NP in all six Wilson coefficients. A more refined analysis is possible by studying the full angular distribution of the latter decays. In this way, it can also be tested whether anomalies in various observables have a consistent NP interpretation.

Global fits of the relevant Wilson coefficients to the data have been performed by a number of groups \cite{Altmannshofer:2017yso,Capdevila:2017bsm,DAmico:2017mtc,Geng:2017svp}. The outcome of this exercise is that a relatively large NP contribution tp the Wilson coefficient $C_9$, 
\begin{equation}
C_9^\text{NP}\simeq -1\,,
\end{equation}
 is required to achieve a significant improvement of the fit with respect to the SM, at the level of $>4\sigma$. Relevant NP contributions to $C'_9$ and/or $C_{10}$ are also allowed by the fit, but not as strictly required.

Before investigating the possible NP interpretations of this anomaly, let us take a look at the hadronic uncertainties in the semileptonic decays $B\to K^{(*)}\mu^+\mu^-$. In the factorisation limit, these decays are well described in terms of $B\to M$ form factors comprising the non-perturbative interactions between the decaying $B$ meson and the final state meson $M=K,K^*$. These form factors can be calculated by lattice QCD \cite{Horgan:2013hoa} and light-cone sum rule \cite{Straub:2015ica} techniques, allowing for further systematic improvements in the near future. For the non-factorisable corrections \cite{Khodjamirian:2010vf,Jager:2012uw,Descotes-Genon:2014uoa,Jager:2014rwa,Lyon:2014hpa}, on the other hand, no systematic theoretical treatment is currently known, and their size can only be estimated. The dominant contribution in this context stems from charm loop effects that are expected to be most relevant in the $q^2$ region below the $c\bar c$ resonances.

A popular approach to deal with the uncertainties from hadronic effects is the construction of observables in which the latter cancel. For the angular distribution of the $B\to K^*\mu^+\mu^-$ final state, the optimised observables $P_i, P'_i$ have been developed \cite{Matias:2012xw,DescotesGenon:2012zf} which are designed to be independent of hadronic form-factors at leading order.  However, one should keep in mind that they are still susceptible to non-factorisable effects \cite{Jager:2012uw,Jager:2014rwa}. To also get rid of the latter uncertainties, the LFU ratios $R_K$, $R_{K^*}$ defined in \eqref{eq:RK} have been suggested \cite{Hiller:2014ula}. They are theoretically extremely clean \cite{Bordone:2016gaq}, as in the SM the only departure from unity is generated by the small muon mass. 
The anomalies at the $2.5\sigma$ level reported by LHCb in the LFU ratios $R_K$ and $R_{K^*}$ therefore deserve particular attention. If experimentally confirmed, the presence of LFU violating NP will be unambiguously proven. However, for the time being one has to await further experimental investigation, like measurements of additional LFU observables and independent confirmations by other experimental collaborations. Luckily, the LHC experiments are currently collecting more data and the first physics run of Belle 2 will start in late 2018, so we should have a definite answer fairly soon.

Interestingly, the various model-independent analyses \cite{Altmannshofer:2017yso,Capdevila:2017bsm,DAmico:2017mtc,Geng:2017svp} show that the anomalies in $R_K$ and $R_{K^*}$ can be resolved by the same NP contribution as the $P'_5$ anomaly, if the NP is assumed to enter only in the muon channel. Due to the large experimental uncertainties, however, also a sizeable NP effect in the electron channel cannot be excluded at present. It should also be noted that the significant suppression of $R_{K^*}$ below the SM prediction also in the very low $q^2$ region, $q^2 \in [0.045,1.1]\,\text{GeV}^2$, calls for a non-zero lepton flavour dependent NP contribution to $C_{10}$ \cite{Geng:2017svp}. Yet, it is impossible to accommodate the experimental central value by means of any NP scenario.

Having established the necessary NP pattern in the effective theory language, it is instructive to consider possible NP models that induce the required contributions. In the most popular NP models, the $b\to s\mu^+\mu^-$ transition is mediated at the tree level, providing a good explanation of the relatively large NP contribution to $C_9$ whose SM contribution is loop-suppressed. Most widely discussed in the literature are variants of an extra neutral $Z'$ gauge boson mediating the flavour changing $b\to s$ transitions, and coupling to muons \cite{Altmannshofer:2013foa,Gauld:2013qba,Gauld:2013qja,Buras:2013qja,Chiang:2016qov,DiChiara:2017cjq}. For example, a model with gauged $L_\mu-L_\tau$ number has been suggested in \cite{Altmannshofer:2014cfa} and subsequently studied in \cite{Altmannshofer:2015mqa,Altmannshofer:2016oaq,Altmannshofer:2016jzy}. In this setup, the flavour changing coupling to the $b\to s$ quark current has been achieved by the presence of extra heavy vectorlike quarks that mix with the SM quarks and are charged under $L_\mu-L_\tau$. The possibility of a $Z'$  resonance of a composite sector has also been investigated \cite{Niehoff:2015bfa,DAmico:2017mtc}. Typically however, in this class of models, a different pattern of NP effects is expected \cite{Altmannshofer:2013foa}. Similar conclusions have also been drawn in the context of Randall-Sundrum models \cite{Blanke:2008yr,Bauer:2009cf,Blanke:2012tv}, which are the 5D dual of a certain type of 4D composite models.

Another popular explanation for the observed anomalies are leptoquark models \cite{Hiller:2014yaa,Gripaios:2014tna,Dorsner:2016wpm}. Similar to the case of the $R(D^{(*)})$ anomaly, also here various realisations in terms of the leptoquark spin and gauge representation are possible. Note that in this class of models, large LFU violating effects are particularly straightforward to accommodate.

While tree-level scenarios are most commonly investigated in the context of the $b\to s$ anomalies, it is also possible to address the problem in terms of loop-induced NP contributions. Box contributions of new particles\cite{Gripaios:2015gra,Arnan:2016cpy} as well as $Z'$ penguins \cite{Kamenik:2017tnu} have been investigated. A $Z'$ model with a loop-induced coupling to muons has been suggested in \cite{Belanger:2015nma}. An explanation in terms of $Z$ penguin effects, as predicted by many popular NP models like the MSSM \cite{Altmannshofer:2013foa} or the Littlest Higgs model with T-parity \cite{Blanke:2006eb,Blanke:2015wba}, on the other hand, is incompatible with the requirement of a large NP contribution to the Wilson coefficient $C_9$. In addition, $Z$ penguin models do not generate any new LFU violating effects.

Important constraints on NP models explaining the $b\to s$ anomalies are obtained from the well-measured $B_s-\bar B_s$ mixing observables \cite{Altmannshofer:2013foa}, the experimental upper bounds on the $B\to K^{(*)}\nu\bar\nu$ decay rates \cite{Buras:2014fpa,Calibbi:2015kma}, as well as the anomalous magnetic moment of the muon, $(g-2)_\mu$ \cite{Bauer:2015knc}. However, they are easier to accommodate than in the case of the $R(D^{(*)})$ anomaly. In addition, the measured SM-like high-$p_T$ dilepton tails cut into the parameter space of NP models explaining the anomalies \cite{Greljo:2017vvb}.

\section{Conclusions and outlook}
\label{sec:sum}

We have provided an overview over the theoretical implications of the  $B$ decay anomalies recently observed by the LHCb collaboration, but also by the $B$ factories BaBar and Belle. Though being short and rather superficial, this summary will hopefully turn out useful in particular to people outside of the community of $B$ physics experts.

The $R(D^{(*)})$ anomaly is experimentally quite well established, yet it calls for an unexpectedly large NP contribution. The known models face substantial constraints from complementary $B$ decay observables, from EW precision measurements and from high-$p_T$ studies, so that a full resolution of the anomaly appears difficult if not impossible.
A NP explanation of the $b\to s$ anomalies including the LFU violating observables $R_{K^{(*)}}$ is easier from that perspective. On the other hand, experimentally these anomalies seem to be on less solid grounds, being driven so far only by the results of one experimental collaboration. If however the LFU anomalies persist, due to their extreme theoretical cleanliness, they will by an unambiguous sign of physics beyond the SM. 

If, in either case, the presence of new LFU violating interactions is confirmed, searches for deviations also in other LFU and lepton flavour violating observables will be crucial in order to identify the underlying lepton flavour breaking effect. 


\end{document}